\documentclass[11pt,a4paper]{article}
\usepackage[utf8]{inputenc}
\usepackage[T1]{fontenc}
\usepackage{lmodern}
\usepackage[margin=1in]{geometry}
\usepackage{microtype}
\usepackage{amsmath,amsfonts,amssymb}
\usepackage[hidelinks]{hyperref}
\usepackage{parskip}
\usepackage{graphicx}
\usepackage{booktabs}
\usepackage{xcolor}
\usepackage{listings}
\usepackage{caption}
\usepackage{natbib}

\definecolor{codebg}{RGB}{245,245,248}
\definecolor{keyword}{RGB}{0,0,150}
\definecolor{comment}{RGB}{0,120,0}
\definecolor{string}{RGB}{150,0,0}

\title{Stratified Bootstrap Test Package}
\author{Ehsan Mohammadi, Fanghua Chen, Yizhou Cai, Yun Yang, Ting Fung Ma, Lu Zhou}
\date{October 17, 2025}

\begin{document}
\maketitle

\begin{abstract}
The Stratified Bootstrap Test (SBT) provides a nonparametric, resampling-based framework for assessing the stability of group-specific ranking patterns in multivariate survey or rating data. By repeatedly resampling observations and examining whether a group’s top-ranked items remain among the highest-scoring categories across bootstrap samples, SBT quantifies ranking robustness through a non-containment index. In parallel, the stratified bootstrap test extends this framework to formal statistical inference by testing ordering hypotheses among population means. Through resampling within groups, the method approximates the null distribution of ranking-based test statistics without relying on distributional assumptions. Together, these techniques enable both descriptive and inferential evaluation of ranking consistency, detection of aberrant or adversarial response patterns, and rigorous comparison of groups in applications such as survey analysis, item response assessment, and fairness auditing in AI systems.
\end{abstract}

\bigskip

\section{Package Introduction}

Recently, the task of testing and cleaning massive datasets that contain mixed variable types has emerged as a central difficulty in statistical inference. Such data typically involve many variables with complex interdependencies, missingness, and computational challenges. Traditional inferential procedures (for example, ANOVA) are often inadequate for determining whether observed component differences are substantive or whether the ordering of group means is robust to sampling variation. To overcome these limitations, we introduce the \emph{Stratified Bootstrap Test} (SBT), a nonparametric resampling procedure designed to evaluate whether an observed ranking of group-level summaries reflects genuine population structure rather than random fluctuation.

The \emph{Stratified Bootstrap Test} (SBT) package include utility functions for converting textual survey responses to numeric scores, computing group-level summary statistics, and conducting stratified resampling to evaluate ranking containment or ordering events. Applications include response validation in human-in-the-loop labeling pipelines, detection of adversarial or low-quality responses, and fairness auditing in AI systems (e.g., auditing visual labeling systems for gender/racial disparities) \citep{Mohammadi2025, Mohammadi2026}.

\section{Background}

Let the sample means from largest to smallest. Given $G$ groups with sample means $\bar{x}_1,\ldots,\bar{x}_G$, suppose we observe
\begin{equation}
    \bar{x}_1 \ge \bar{x}_2 \ge \cdots \ge \bar{x}_G .
\end{equation}

This observed ordering may arise due to sampling randomness and need not reflect the ordering of the true population means $\mu_1,\ldots,\mu_G$.

The core inferential question is whether the observed ordering is likely under random sampling, or whether it reflects genuine differences in population means. To make this question testable we reformulate it as hypothesis testing for ordering constraints. For a pre-specified split $g\in\{1,\dots,G-1\}$, consider
\begin{equation}
\begin{aligned}
    H_0:\ \min_{i=1,\dots,g}\mu_i &\le \max_{j=g+1,\dots,G}\mu_j\\
ver&sus\\
H_1:\ \min_{i=1,\dots,g}\mu_i &> \max_{j=g+1,\dots,G}\mu_j.
\end{aligned}
\end{equation}

In other settings, one may wish to test a strict total ordering:
\begin{equation}
    H_1^{(\text{total})}:\ \mu_1 > \mu_2 > \cdots > \mu_G,
\end{equation}

with the null hypothesis \(H_0^{(\text{total})}\) being the complement (i.e., at least one inequality fails).

Because these hypotheses concern order structure rather than scalar differences alone, we employ stratified bootstrap (resampling within groups) to approximate the null distribution of ranking-based test statistics without parametric assumptions \citep{Zhang2023, Nichols2001}.

\section{Statistical principle}
Let \(X=(x_{rj})\) denote the observed response matrix with \(n\) rows (respondents) and \(k\) columns (items). Denote by \(I_g\) the index set of respondents in group \(g\) and let \(|I_g|\) be its cardinality.

\subsection{Group means}
For each group \(g\) and item \(j\), compute the group mean:
\[
\bar{x}_{g,j} = \frac{1}{|I_g|}\sum_{r\in I_g} x_{rj}.
\]

\subsection{Top--\(i\) set (containment perspective)}
Sort \(\bar{x}_{g,\cdot}\) in descending order and denote the indices of the top \(i\) items by
\[
T_{g,i} = \mathrm{Top}_i(\bar{x}_{g,1},\dots,\bar{x}_{g,k}).
\]

\subsection{Stratified resampling and event probability}
Conduct stratified resampling within each group: for each group \(i\) resample \(|I_i|\) observations \emph{with replacement} to create a bootstrap/resampled dataset. For resample \(b=1,\dots,B\) compute the group means \(\bar{x}_1^{*(b)},\ldots,\bar{x}_G^{*(b)}\). Define the event of interest for a split \(g\):
\[
E^{*(b)} = \left\{ \min_{i=1,\dots,g} \bar{x}_i^{*(b)} > \max_{j=g+1,\dots,G} \bar{x}_j^{*(b)} \right\}.
\]

Estimate the Monte Carlo probability of this event by
\[
\hat{p} = \frac{1}{B}\sum_{b=1}^B \mathbf{1}\!\left( E^{*(b)} \right),
\]
where \(\mathbf{1}(\cdot)\) is the indicator function. In our recommended settings we use a large \(B\) (e.g. \(B=10{,}000\)) to stabilize the Monte Carlo approximation.

\subsection{Extension to strict total ordering}
In some applications, the goal is to assess whether the population means follow a strict total order consistent with the observed ranking. In this case, we consider
\[
H_1^{(\text{total})}:\ \mu_1 > \mu_2 > \cdots > \mu_G,
\]
with the corresponding null hypothesis given by its complement,
\[
H_0^{(\text{total})}:\ \text{at least one inequality in } \mu_1 > \mu_2 > \cdots > \mu_G \text{ fails.}
\]

Since this statistical hypothesis setting is non-standard, we adopt the Bootstrap test technique
(see \citet{Zhang2023} for a review) by resampling the observations within the groups
(stratification within the group). The Bootstrap test is a general approach for approximating
the distribution of a test statistic (for our case, the ordering of the sample means) under the
null hypothesis without assuming a statistical model. In other words, the bootstrap test is
a non-parametric technique \citep{Nichols2001}. The testing procedure usually relies on
approximating the distribution of the test statistic via some resampling procedure through computer
simulations for $B$ realizations of the test statistic. We calculate the observed test statistic using the
resampled data for each of the $B$ realizations. The statistical significance of the test can then be
assessed by the likelihood of the event among the $B$ realizations of the test statistic. See \citet{Edgington2025} for further discussion.

To retain the grouping structure of the dataset, we resample $n_i$ observations by sampling with
replacement within the $i$-th group. After the resampling step, we compute the corresponding group
sample means. In this setting, our focus is on assessing whether the population means of the first
$g$ groups are all strictly larger than those of the remaining $G-g$ groups. Mathematically, this
corresponds to the event
\[
\min_{i=1,\dots,g} \mu_i > \max_{j=g+1,\dots,G} \mu_j .
\]

The likelihood of this event is estimated by the proportion of
\[
\min_{i=1,\dots,g} \bar{x}_i^{*(b)} > \max_{j=g+1,\dots,G} \bar{x}_j^{*(b)}
\]
among the $B$ simulations, where $\bar{x}_k^{*(b)}$ is the sample mean for the $k$-th group using the $b$-th resampled
dataset.

The resulting Monte Carlo $p$-value is computed as
\[
\hat{p} \;=\; \frac{1}{B}\sum_{b=1}^B \mathbf{1}\!\left( \min_{i=1,\dots,g} \bar{x}_i^{*(b)} > \max_{j=g+1,\dots,G} \bar{x}_j^{*(b)} \right),
\]
where $\mathbf{1}(\cdot)$ is the indicator function that equals 1 if the smallest resampled mean among the first $g$
groups exceeds the largest resampled mean among the remaining $G - g$ groups, and 0 otherwise. In our analysis, we set $B = 10{,}000$ to ensure numerical stability of the Monte Carlo approximation.
\cite{Mohammadi2025, Mohammadi2026}
\section{Implemented functions}
This package includes two primary R functions documented below.

\subsection{\texttt{SingleStratifiedBootstrap()}}
\paragraph{Purpose} For a numeric matrix (rows = observations, columns = items), compute how often a target set of columns (top--\(k\) indices) fails to be fully contained in the stratified bootstrap top--\(k\) sets across \(B\) resamples.

\paragraph{Signature}
\begin{verbatim}
SingleStratifiedBootstrap(
  data,
  n_boot = 1000,
  target_indices,
  summary_fun = mean,
  sample_size = NULL,
  replace = TRUE,
  decreasing = TRUE,
  na.rm = TRUE,
  seed = NULL,
  n_cores = 1L
)
\end{verbatim}

\paragraph{Return} Numeric scalar in \([0,1]\): the non-containment rate,
\[
1 - \frac{\#\{\text{resamples containing target}\}}{n_{\text{boot}}}.
\]

\subsection{\texttt{GetSBT()}}
\paragraph{Purpose} For a given list of group levels and a response matrix, compute:
\begin{itemize}
  \item group-level mean table,
  \item SBT (non-containment) matrix where each column corresponds to \(\text{top}_i\), \(i=1,\dots,k\).
\end{itemize}

\paragraph{Signature}
\begin{verbatim}
GetSBT(
  group_levels,
  group_data,
  response,
  n_boot = 1000,
  response_type = c("likert", "binary", "numeric"),
  likert_map = NULL,
  summary_fun = mean,
  sample_size = NULL,
  replace = TRUE,
  decreasing = TRUE,
  na.rm = TRUE,
  seed = NULL,
  n_cores = 1L,
  min_group_size = 3L
)
\end{verbatim}

\paragraph{Return} A list with:
\begin{itemize}
  \item \texttt{MeanTable}: data.frame (G $\times$ k), group means or proportions.
  \item \texttt{noncontainment}: data.frame (G $\times$ k) where column \texttt{top\_i} is the non-containment rate for top--\(i\).
\end{itemize}

\section{Usage example}
Below is a runnable R example. Use it in an R session or incorporate into a vignette.

\begin{lstlisting}[caption={R example: run SBT on simulated data}]
# Simulate Likert-style data
set.seed(42)
n <- 100
response <- data.frame(
  Q1 = sample(1:5, n, TRUE),
  Q2 = sample(1:5, n, TRUE),
  Q3 = sample(1:5, n, TRUE),
  Q4 = sample(1:5, n, TRUE),
  Q5 = sample(1:5, n, TRUE)
)

# Group labels
group_data <- sample(c("Woman", "Man"), n, TRUE)

# Run GetSBT (assumes functions are loaded)
Result <- GetSBT(
  group_levels = c("Woman", "Man"),
  group_data = group_data,
  response = response,
  n_boot = 500,
  response_type = "likert",
  seed = 123
)

# Inspect outputs
print(Result$MeanTable)         # group means per question
print(Result$noncontainment)    # instability indices top_1 ... top_5
\end{lstlisting}

\section{Interpretation guidance}
\begin{itemize}
  \item \textbf{MeanTable}: For Likert responses, values typically range from 1 to 5. For binary responses, values are proportions in [0,1].
  \item \textbf{noncontainment (SBT)}: Each value is in [0,1]. Low values (near 0) mean the group's top--\(i\) is consistently reproduced by resampling; high values (near 1) mean the top--\(i\) is unstable.
  \item Inspect group sizes; small groups yield noisy resampling estimates. Use the \texttt{min\_group\_size} parameter to control warnings.
  \item To compare groups visually, consider heatmaps of the \texttt{noncontainment} table or bar plots for \texttt{top\_1}.
\end{itemize}

\section{Practical notes and recommendations}
\begin{itemize}
  \item Use named (keyword) arguments when calling \texttt{GetSBT()} to avoid positional argument mismatches if the function signature evolves.
  \item Set \texttt{seed} for reproducibility, particularly if running in parallel.
  \item For large datasets and high \texttt{n\_boot}, use \texttt{n\_cores} (non-Windows) or a cross-platform parallel backend (e.g., \texttt{future.apply}).
  \item If responses are textual, clean common variants (case differences and trailing spaces) or provide a \texttt{likert\_map} to control mapping.
  \item Document group sizes in results; very small groups (\textless 10) can be unstable.
\end{itemize}

\bibliographystyle{plainnat}
\bibliography{document_ref}

@article{Mohammadi2025,
  title = {Who is a scientist? Gender and racial biases in google vision AI},
  volume = {5},
  ISSN = {2730-5961},
  url = {http://dx.doi.org/10.1007/s43681-025-00742-4},
  DOI = {10.1007/s43681-025-00742-4},
  number = {5},
  journal = {AI and Ethics},
  publisher = {Springer Science and Business Media LLC},
  author = {Mohammadi,  Ehsan and Cai,  Yizhou and Novin,  Alamir and Vera,  Valerie and Soltanmohammadi,  Ehsan},
  year = {2025},
  month = may,
  pages = {4993–5010}
}

@article{Mohammadi2026,
  title = {Is generative AI reshaping academic practices worldwide? A survey of adoption,  benefits,  and concerns},
  volume = {63},
  ISSN = {0306-4573},
  url = {http://dx.doi.org/10.1016/j.ipm.2025.104350},
  DOI = {10.1016/j.ipm.2025.104350},
  number = {1},
  journal = {Information Processing \&amp; Management},
  publisher = {Elsevier BV},
  author = {Mohammadi,  Ehsan and Thelwall,  Mike and Cai,  Yizhou and Collier,  Taylor and Tahamtan,  Iman and Eftekhar,  Azar},
  year = {2026},
  month = jan,
  pages = {104350}
}

@article{Nichols2001,
  title = {Nonparametric permutation tests for functional neuroimaging: A primer with examples},
  volume = {15},
  ISSN = {1097-0193},
  url = {http://dx.doi.org/10.1002/hbm.1058},
  DOI = {10.1002/hbm.1058},
  number = {1},
  journal = {Human Brain Mapping},
  publisher = {Wiley},
  author = {Nichols,  Thomas E. and Holmes,  Andrew P.},
  year = {2001},
  month = oct,
  pages = {1–25}
}

@article{Zhang2023,
  title = {What is a Randomization Test?},
  volume = {118},
  ISSN = {1537-274X},
  url = {http://dx.doi.org/10.1080/01621459.2023.2199814},
  DOI = {10.1080/01621459.2023.2199814},
  number = {544},
  journal = {Journal of the American Statistical Association},
  publisher = {Informa UK Limited},
  author = {Zhang,  Yao and Zhao,  Qingyuan},
  year = {2023},
  month = may,
  pages = {2928–2942}
}

@inbook{Edgington2025,
  title = {Randomization Tests},
  ISBN = {9783662693599},
  url = {http://dx.doi.org/10.1007/978-3-662-69359-9_510},
  DOI = {10.1007/978-3-662-69359-9_510},
  booktitle = {International Encyclopedia of Statistical Science},
  publisher = {Springer Berlin Heidelberg},
  author = {Edgington,  Eugene S.},
  year = {2025},
  pages = {2095–2096}
}

\end{document}